\documentclass{ws-p8-50x6-00}

\begin{document}

\title{Electroweak properties of the nucleon \\
in a chiral constituent quark model}

\author{S. Boffi and \underline{M. Radici}}

\address{Dipartimento di Fisica Nucleare e Teorica, 
Universit\`a di Pavia, and \\
INFN, Sezione di Pavia, I-27100 Pavia, Italy}

\author{L. Glozman, W. Plessas and R.F. Wagenbrunn}

\address{Institut f\"ur Theoretische Physik, 
Universit\"at Graz, A-8010 Graz, Austria}  

\author{W. Klink}

\address{Department of Physics and Astronomy, 
University of Iowa, Iowa City, IA 52242, USA}  

\maketitle

\abstracts{Results for all elastic electroweak 
nucleon form factors are presented for the 
chiral constituent quark model based on 
Goldstone-boson-exchange dynamics. The calculations 
are performed in a covariant framework using
the point-form approach to relativistic quantum mechanics.
The direct predictions of the model yield
a remarkably consistent picture of the electroweak 
nucleon structure.}

\begin{figure}[th]
\begin{center}
\epsfxsize=16pc 
\epsfbox{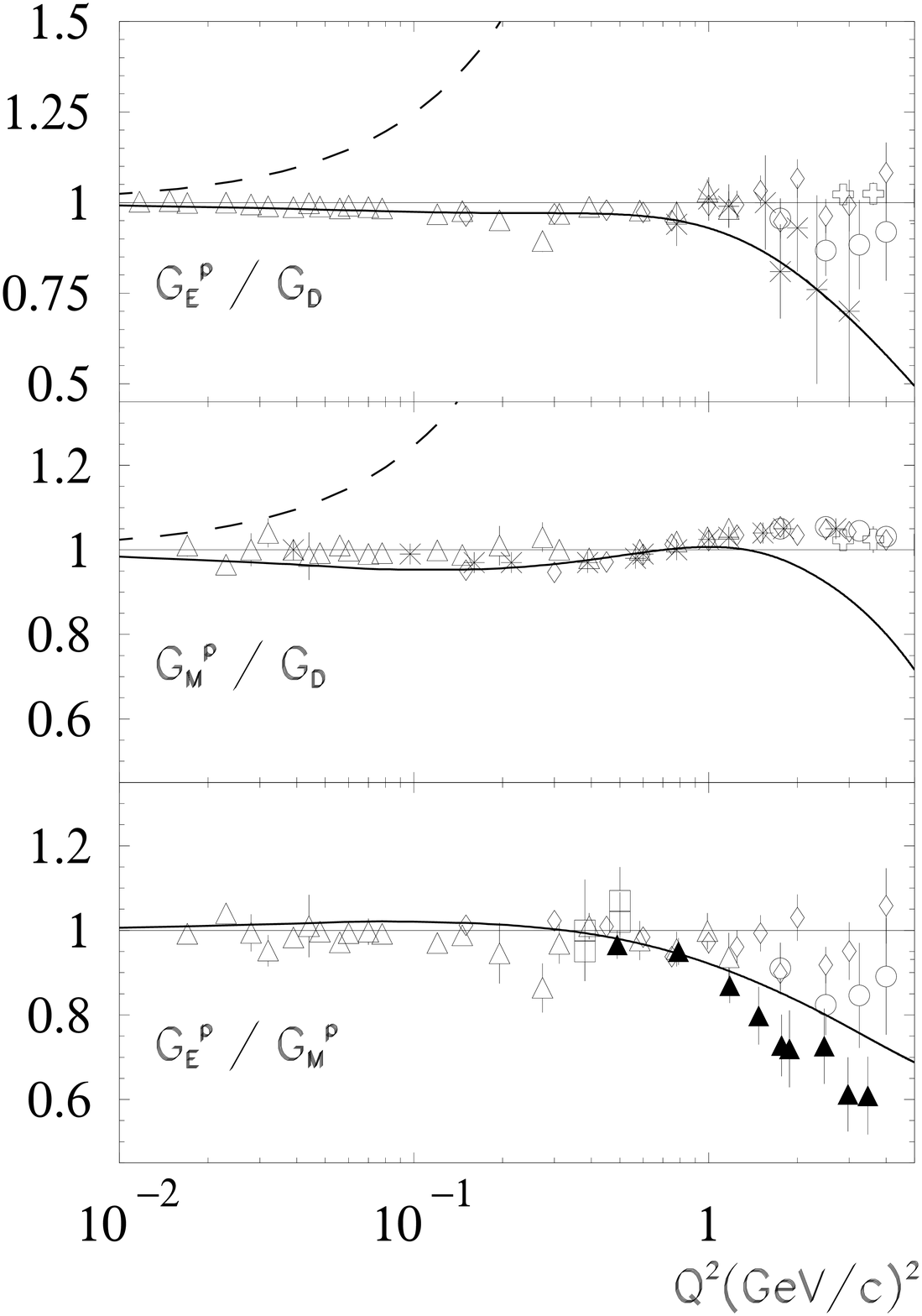} 
\vspace{-.3truecm}
\caption{Proton electric and magnetic form factors. Top and middle panels: 
ratios of electric ($G_E^p$) and magnetic ($G_M^p$) proton form factors to 
the standard dipole parametrization $G_D$. Bottom panel: ratio of 
$G_E^p$ to $G_M^p$. PFSA predictions of the GBE CQM (solid lines) are 
compared with nonrelativistic results (dashed lines) and experiment. 
In the top and middle panels the experimental data are from 
Ref.~{\protect\cite{proton}}. In the bottom panel recent data from 
TJNAF~{\protect\cite{jones}} (filled triangles) are shown together 
with various older data points (see Ref.~{\protect\cite{jones}} 
and references therein). All the ratios are normalized to 1 at 
$Q^2=0$. \label{fig:fig1}}
\end{center}
\end{figure}

\begin{figure}[h]
\begin{center}
\epsfxsize=25pc 
\epsfbox{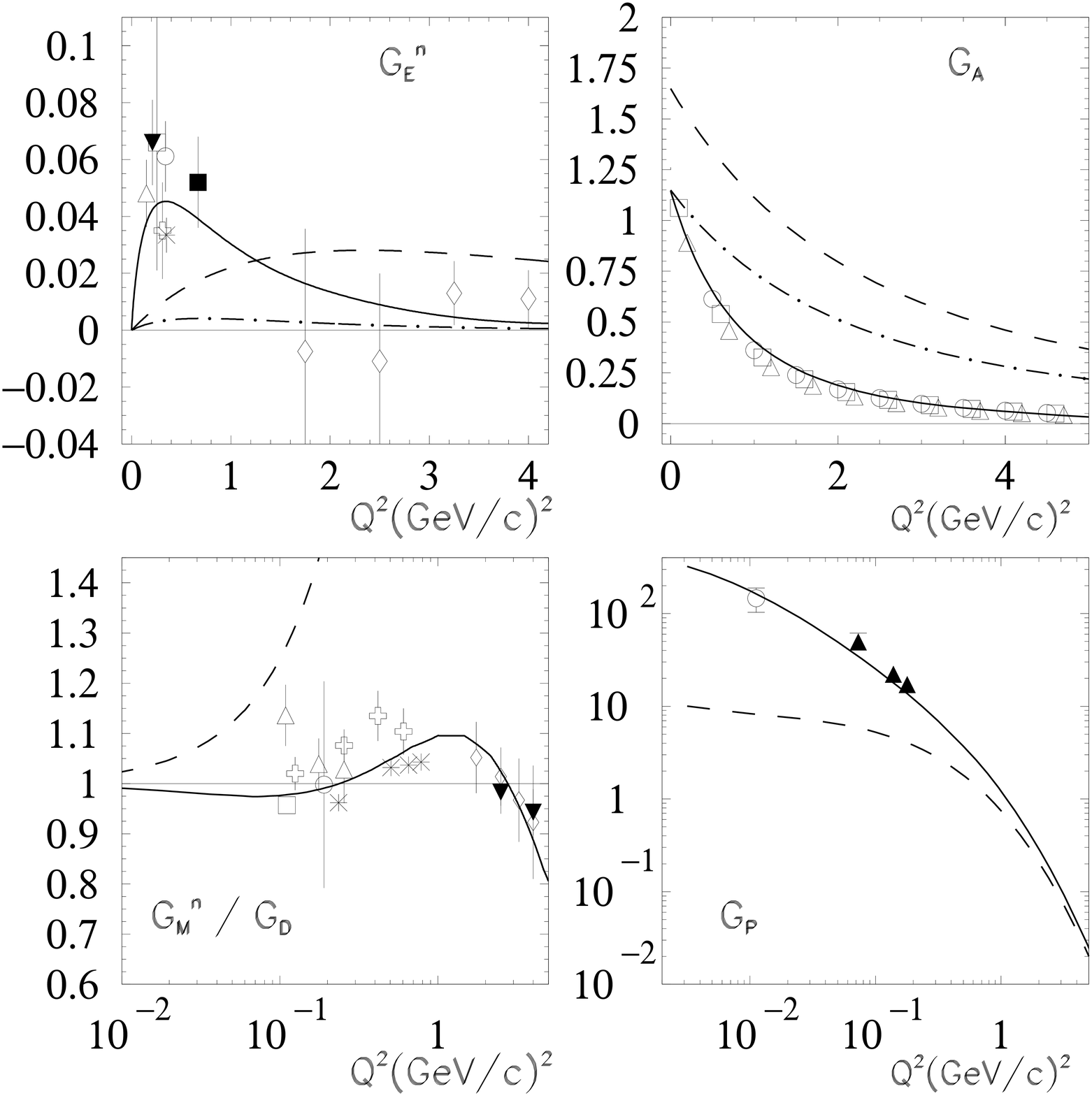}
\caption{Left panel: neutron electric and magnetic form factors; in the top  
panel, $G_E^n$; in the bottom panel, ratio of $G_M^n$ to the standard dipole  
parametrization $G_D$, normalized to 1 at $Q^2=0$; solid and dashed
lines as in Fig.~{\protect\ref{fig:fig1}}; the dot-dashed line 
represents the PFSA results for the case with confinement only; 
experimental data are from Ref.~{\protect\cite{gen}} (top) and
Ref.~{\protect\cite{gmn}} (bottom). 
Right panel: nucleon axial and induced pseudoscalar form factors $G_A$ and
$G_P$, respectively; the PFSA predictions of the GBE CQM are always
represented by solid lines; in the top panel, a comparison is given to the 
nonrelativistic results (dashed) and to the case with a relativistic current
operator but no boosts included (dot-dashed); experimental data 
are shown assuming a dipole parameterization with the axial 
mass value $M_A$ deduced from pion electroproduction (world average: 
squares, Mainz experiment~{\protect\cite{mainz}}: circles) and from neutrino
scattering~{\protect\cite{neut}} (triangles); in the bottom panel, the dashed
line refers to the calculation of $G_P$ without any pion-pole contribution;   
the experimental data are from Ref.~{\protect\cite{gp-exp}}. \label{fig:fig2}}
\vspace{-.6truecm}
\end{center}
\end{figure}

We present results for the elastic electroweak nucleon 
observables as a progress report of a more comprehensive 
programme aiming at a consistent description of the 
electroweak properties of baryons at low energy. The theoretical context 
is represented by the chiral Constituent Quark Model (CQM) based 
on the Goldstone-boson exchange (GBE) quark-quark interaction,  
that is induced by the spontaneous chiral symmetry breaking in QCD 
and that accurately reproduces the baryon spectrum of light and strange
flavors~\cite{gbe}. The dynamics of quarks inside the nucleon 
is essentially relativistic. Therefore, we have adopted the 
point-form realization of relativistic quantum mechanics, where 
the boost generators are interaction-free and make the theory 
manifestly covariant~\cite{klink}. The electromagnetic photon-quark 
interaction is assumed point-like, but in point-form the momentum 
delivered to the nucleon is different from the one delivered to the 
struck quark; hence, we will name this approach the Point-form Spectator 
Approximation (PFSA)~\cite{grazpv-ps}. The quark wave functions deduced from 
fitting the baryon spectrum are used as input and no further parameter is 
introduced, since quarks are considered point-like and the point-form 
allows for an exact calculation of all boosts required by a covariant 
description. Results have recently been published for 
electromagnetic~\cite{grazpv-em}, axial~\cite{grazpv-ax} and 
pseudoscalar~\cite{grazpv-ps} nucleon form factors. They are summarized here 
in Figs.~\ref{fig:fig1}-\ref{fig:fig2}, 
and in Tab.~\ref{tab:tabI}. 

\begin{table}[h]
\caption{Proton and neutron charge radii and magnetic moments 
as well as nucleon axial radius and axial charge.
Predictions of the GBE CQM in PFSA (third column), in nonrelativistic 
approximation (NRIA, fourth column), and with the confinement interaction 
only (last column).\label{tab:tabI}}
\begin{center}
\footnotesize
\begin{tabular}{|c|c|c|c|c|}
\hline
{} & Exp. & PFSA & NRIA & Conf. \\
\hline 
$r_p^2$ [fm$^2$] & 0.780(25)~\cite{rp} & 0.81 & 0.10 & 0.37  \\
$r_n^2$ [fm$^2$] & -0.113(7)~\cite{rn} & -0.13 & -0.01 & -0.01 \\
$\mu_p$ [n.m.] & 2.792847337(29)~\cite{pdg} & 2.7 & 2.74 & 1.84 \\
$\mu_n$ [n.m.] & -1.91304270(5)~\cite{pdg} & -1.7 & -1.82 & -1.20 \\
$<r_A^2>^{\frac{1}{2}}$ [fm]& 0.635(23)~\cite{mainz} & 0.53 & 0.36 & 0.43 \\
$g_A$ & 1.255$\pm$ 0.006~\cite{pdg} & 1.15 & 1.65 & 1.29 \\
\hline
\end{tabular}
\vspace{-.5truecm}
\end{center}
\end{table}

The agreement with experimental data is 
remarkable and it indicates that by a proper choice of low-energy degrees 
of freedom a quark model is capable of describing the spectroscopy and the 
low-energy dynamics of baryons at the same time. 
However, a more detailed comparison with data shows that there is still 
room for quantitative improvements, e.g. by considering two-body 
electromagnetic current operators and constituent quark sizes.

\end{document}